\documentstyle[11pt,aaspp4]{article}

\begin{document}

\title{Physical Mechanisms for the Variable Spin-down of SGR~$1900+14$}

\author{
Christopher~Thompson\altaffilmark{1},
Robert~C.~Duncan\altaffilmark{2}, 
Peter~M.~Woods\altaffilmark{3,5},
Chryssa~Kouveliotou\altaffilmark{4,5},
Mark~H.~Finger\altaffilmark{4,5}, and
Jan~van~Paradijs\altaffilmark{3,6}
}

\altaffiltext{1}{Department of Physics and Astronomy, University of North
Carolina, Philips Hall, Chapel Hill, NC 27599-3255}
\altaffiltext{2}{Department of Astronomy, University of Texas, RLM 15.308,
Austin, TX 78712-1083}
\altaffiltext{3}{Department of Physics, University of Alabama in Huntsville, 
Huntsville, AL 35899}
\altaffiltext{4}{Universities Space Research Association}
\altaffiltext{5}{NASA Marshall Space Flight Center, SD50, Huntsville, AL
35812}
\altaffiltext{6}{Astronomical Institute ``Anton Pannekoek'', University of 
Amsterdam, 403 Kruislaan, 1098 SJ Amsterdam, NL}

\def\be{\begin{equation}}
\def\ee{\end{equation}}
\begin{abstract}

We consider the physical implications of the rapid spindown of Soft
Gamma Repeater 1900+14 reported by Woods et al.  
During an 80 day interval between June 1998 and the large outburst on
August 27 1998, the mean spin-down rate increased by a factor 2.3,
resulting in a positive period offset of $\Delta P/P = 1\times 10^{-4}$. 
A radiation-hydrodynamical outflow associated with the August 27th event could
impart the required torque, but only if the dipole magnetic field is stronger
than $\sim 10^{14}$ G and the outflow lasts longer and/or is more energetic
than the observed X-ray flare. A positive period increment is also a
natural consequence of a gradual, plastic deformation of the neutron star
crust by an intense magnetic field, which forces the neutron
superfluid to rotate more slowly than the crust.  Sudden unpinning of the
neutron vortex lines during the August 27th event would then induce a glitch
opposite in sign to those observed in young pulsars, but of a much larger
magnitude as a result of the slower rotation.

The change in the persistent X-ray lightcurve following the August 27
event is ascribed to continued particle heating in the active region of
that outburst.  The enhanced X-ray output can be powered by a steady
current flowing through the magnetosphere, induced by the twisting motion of
the crust.  The long term rate of spindown appears to be accelerated with
respect to a simple magnetic dipole torque.  Accelerated spindown of a 
seismically-active magnetar will occur when its persistent output of
Alfv\'en waves and particles exceeds its spindown luminosity.  We suggest
that SGRs experience some episodes of relative inactivity, with diminished
$\dot{P}$, and that such inactive magnetars are observed as Anomalous
X-ray Pulsars (AXPs).   The reappearence of persistent X-ray emission
from SGR 1900+14 within one day of the August 27 event provides strong
evidence that the persistent emission is not powered by accretion.

\end{abstract}


\section{Introduction}

Woods et al.~(1999c; hereafter Paper I) have shown that over the period
September 1996 -- May 1999, the spin-down history of SGR~$1900+14$ is
generally smooth, with an average rate of 
6 $\times$ 10$^{-11}$ s s$^{-1}$.  However, during an 80 day
interval starting in June 1998 which contains the extremely energetic August
27 flare (Hurley et al.~1999a; Mazets et al.~1999), the average
spindown rate of SGR~$1900+14$ increased by a factor $\sim$ 2.3. 
The sampling of the period history of SGR~$1900+14$ is insufficient to
distinguish between a long-term (i.e. 80 days) increase of the spin-down rate
to an enhanced value and a sudden increase (a `braking' glitch) in the spin
period connected with the luminous August 27 flare.

In this paper, we investigate several physical processes that may
generate a positive period increment of the observed magnitude ($\Delta P/P
\sim 10^{-4}$) directly associated with the August 27 flare.  We
focus on two mechanisms:  a  particle wind coinciding with the period of
hyper-Eddington radiative flux; and an exchange of angular momentum between the
crustal neutron superfluid and the rest of the neutron star.  We show that both
models point to the presence of an intense magnetic field.  The
change in the persistent pulse profile of SGR 1900+14 following the
August 27 outburst is considered, and related to continuing particle
output in the active region of the burst.  We also
consider mechanisms that could drive the (nearly) steady spindown observed 
in both SGRs and AXPs, as well as departures from uniform spindown.

\section{Braking driven by a particle outflow}

The radiative flux during the oscillatory tail of the August 27 event decreased
from $1\times 10^{42}\,(D/10~{\rm kpc})^2$ erg/s, with an exponential time
constant of $\sim 90$ s (Mazets et al.~1999). The net energy in the tail,
radiated in photons of energy $>$ 15 keV, was
$\sim 5\times 10^{43}(D/10~{\rm kpc})^2$ erg. 
The tail was preceded by much
harder, narrow pulse of duration $\sim  0.35$ s and energy $> 7\times
10^{43}(D/10~{\rm kpc})^2$ erg (Mazets et al.~1999). The very fast rise 
time of
$\sim 10^{-3}$ s points convincingly to an energy source internal to the
neutron star.  Just as in the case of the 1979 March 5 event, several arguments
indicate the presence of a magnetic field stronger than $10^{14}$ G (Thompson
\& Duncan 1995; hereafter ``TD95").  Not only can such a field spin down 
the star to its observed
5.16 s period (Hurley et al.~1999c; Kouveliotou et al.~1999), but it  can power
the burst by inducing a large-scale fracture of the neutron star crust. 
Indeed, only a fraction $\sim 10^{-2}\,(B_\star/10\,B_{\rm QED})^{-2}$ of the
external dipole magnetic energy must be tapped, 
where $B_{\rm QED} \equiv 4.4\times 10^{13}$ G. \  This allows for individual SGR
sources to emit $\gtrsim 10^2$ such giant flares over their 
$\sim 10^4$ yr active
lifetimes.    More generally, {\it any} energy source that
excites internal seismic modes of the neutron star must be  combined with a
magnetic field of this strength, if seismic energy is to be transported across
the stellar surface at the (minimum) rate observed in the initial spike (cf.
Blaes et al.~1989). A field stronger than $1.5\times
10^{14}\,(E/6\times10^{43}~{\rm erg})^{1/2}\, (\Delta R/{\rm
10~km})^{-3/2}[(1+\Delta R/R_\star)/2]^{3}$ G is also required to confine the
energy radiated in the oscillatory tail (Hurley et al.~1999a), which maintained
a very constant temperature even while the radiative flux  declined by an order
of magnitude (Mazets et al.~1999).

The radiative flux was high enough throughout the August 27 event
to advect outward a large amount of baryonic plasma at relativistic
speed.  Even though one photon polarization mode (the E-mode) has a suppressed
scattering cross-section when $B> B_{\rm QED}$ (Paczy\'nski
1992), splitting of E-mode photons will regenerate the O-mode outside
the E-mode scattering photosphere, and ensure than the radiation and
matter are hydrodynamically coupled near the stellar surface (TD95).  
Matter will continue to accumulate further
out in the magnetosphere during the burst, but cannot exceed 
$\tau_{\rm T} \sim 1$ outside
a radius where the energy density of the freely streaming photons
exceeds the dipole magnetic energy density,
\be
{L_{\rm X}\over 4\pi R_{\rm A}^2c} \sim {B_\star^2\over 4\pi}\left({R_{\rm A}\over
R_\star}\right)^{-6},
\ee
or equivalently
\be\label{Ralf}
{R_{\rm A}\over R_\star} \sim \left({B_\star^2 R_\star^2c\over L_{\rm X}}\right)^{1/4}
= 280\,\left({B_\star\over 10~B_{\rm QED}}\right)^{1/2}\,
\left({\Delta E_{\rm X}\over 10^{44}~{\rm erg}}\right)^{-1/4}\,
\left({\Delta t_{\rm burst}\over 100~{\rm s}}\right)^{1/4}.
\ee
The radiation pressure acting on the suspended matter will overcome
the dipole magnetic pressure at a radius $\leq R_{\rm A}$; the same is true
for the ram pressure of matter streaming relativistically outward along the
dipole field lines.

Photons scattering last at radius $R_{\rm A}$ and polar angle $\theta$ (or
relativistic matter escaping the dipole magnetic field from the same
position) will carry a specific angular momentum 
$\sim \Omega R_{\rm A}^2\sin^2\theta$.  The net loss of angular momentum
corresponding to an energy release $\Delta E$ is
\be\label{domega}
I_\star\Delta\Omega \simeq -{\Delta E\over c^2}\Omega R_{\rm A}^2\sin^2\theta.
\ee
The period increase accumulated on a timescale $\Delta t_{\rm burst}$ is
largest if the outflow is concentrated in the equatorial plane of the star:
\be\label{deltap}
{\Delta P\over P} \simeq (\Delta E\Delta t_{\rm burst})^{1/2}\,
{B_\star R_\star^3\over I_\star c^{3/2}} =
8\times 10^{-6}\,\left({\Delta E\over 10^{44}~{\rm erg}}\right)^{1/2}
\,\left({\Delta t_{\rm burst}\over 100~{\rm s}}\right)^{1/2}\,
\left({B_\star\over 10~B_{\rm QED}}\right).
\ee

The torque is negligible if the dipole field is in the range
$B_\star \sim 0.1 B_{\rm QED}$ typical of ordinary radio pulsars. 
Even for $B_\star \sim 10~B_{\rm QED}$ this mechanism 
can induce $\Delta P/P \sim 1\times 10^{-4}$ only if the the outflow
lasts longer than the observed duration of the oscillatory tail.
Release of $\sim 10^{44}$ erg over $\sim 10^4$ s would suffice;
but extending the duration of the outflow to $\sim 10^5$ s would imply
$\dot P \sim 1.3\times 10^{-8}$ one day after the August 27
event, in contradiction with the measured value $200$ times smaller.  
Note also that the short initial spike is expected to impart a negligible
torque to the star.
This is the basic reason that {\it persistent} fluxes of Alfv\'en waves
and particles are more effective at spinning down a magnetar than
are sudden, short bursts of equal fluence.  

One might consider increasing the torque by increasing the inertia of
the outflow, so that it moves subrelativistically at the Alfv\'en 
surface, at speed $V$. \  For a fixed kinetic luminosity, 
$\dot E = (1/2) \dot{M} V^2$,  the Alfv\'en radius scales in 
proportion to $(V/c)^{1/4}$, and one finds
\be
{\Delta P\over P} \simeq 1 \times 10^{-4} \, 
\left({\Delta E\over 10^{44}~{\rm erg}}\right)^{1/2}
\,\left({\Delta t_{\rm burst}\over 100~{\rm s}}\right)^{1/2}\,
\left({B_\star\over 10~B_{\rm QED}}\right)
\,\left({V\over 0.2 c}\right)^{-3/2} .
\ee
However, the energy needed to lift this material from the surface
of the neutron star exceeds $\Delta E = \int \dot E dt$ by a factor
$\sim 10\,(V/0.2~c)^{-2}$ (assuming $GM_\star/(R_\star c^2) = 0.2$).
This scenario therefore requires some fine-tuning, if the flow is
to remain subrelativistic far from the neutron star.

Moreover, such a slow outflow is very thick to Thomson scattering and
free-free absorption.  The Thomson depth along a radial line through
the outflow is
\be
\tau_{\rm T}(R_{\rm A}) = 10\,
\left({\Delta E\over 10^{44}~{\rm erg}}\right)^{5/4}\,
\left({B_\star\over 10~B_{\rm QED}}\right)^{-1/2}\,\left({\Delta t_{\rm burst}\over
100~{\rm s}}\right)^{-5/4}\,\left({V\over c}\right)^{-13/4}
\ee
at the Alfv\'en radius.  The free-free optical depth is
\be
\tau_{\rm ff} \simeq {\alpha_{\rm em}\bar g_{\rm ff}\over 3^{1/2}(2\pi)^{3/2}}
\left({kT\over m_ec^2}\right)^{-1/2}\,
{\tau_{\rm T}^2 (hc)^3\over \sigma_{\rm T} R (kT)^3}\,\;f\left({h\nu\over kT}\right),
\ee
where
\be
f\left({h\nu\over kT}\right) \equiv \left({h\nu\over kT}\right)^{-3}\,
\left(1-e^{-h\nu/kT}\right),
\ee
and $\alpha_{\rm em} = 1/137$ is the fine structure constant.
This becomes
\be
\tau_{\rm ff}(R) = 3\times 10^{-2}\,\left({R\over R_{\rm A}}\right)^{-5}\,
\left({\Delta P/P\over 10^{-4}}\right)^{5/4}\,
\left({\Delta E\over 10^{44}~{\rm erg}}\right)^{1/2}\,
\left({B_\star\over 10~B_{\rm QED}}\right)^{-16/3}\,
\left({\Delta t_{\rm burst}\over 100~{\rm s}}\right)^{-5}\,
f\left({h\nu\over kT}\right).
\ee
Here, we have substituted the value of $V/c$ needed to generate
the observed $\Delta P/P$.  
Notice that the magnetic dipole field and burst duration enter 
into $\tau_{\rm ff}$ with strong negative powers.  
The optical depth through a flow along rigid dipole magnetic field
lines is $\tau_{\rm T}(R) = (R/R_{\rm A})^{-2}\,\tau_{\rm T}(R_{\rm A})$ at constant $V$.

This calculation indicates that the
flow will be degraded to a black body temperature corresponding
to an emission radius of $\sim 100\,R_\star$ = 1000 km, which
is $\sim 1$ keV at a luminosity $\sim 10^4\,L_{\rm edd}$, far below
the observed value (Mazets et al.~1999; Feroci et al.~1999).
Note, however, that Inan et al.~(1999) found evidence 
for an intense ionizing flux of soft X-rays in the Earth's ionosphere,
coincident with the first second of the August 27th event.   They
fit this ionization data with an incident spectrum containing
two thermal components, of temperatures 200 and 5 keV, and with the soft
component carrying $80\%$ of the energy flux at 5 keV.   This model
contrasts with the initial spectrum of the August 27 event measured by 
BeppoSAX, which contained a very hard power-law component 
($\nu F_\nu \propto \nu^{1/2}$:  Feroci et al.~1999).  The effects
of pair creation on the ionization rate have yet to be quantified.

The four-pronged profile seen within the later pulses of the August 27 event
(Feroci et al.~1999; Mazets et al.~1999) has a plausible interpretation
in the magnetar model.  The radiation-hydrodynamical outflow originates
near the surface of the neutron star, where the opacity of X-ray photons
moving across the magnetic field lines is smallest (TD95).
This is the case even if the trapped $e^\pm$ fireball that powers the
burst extends well beyond the stellar surface.   In this model, the
pattern of the emergent X-ray flux is a convolution of the multipolar
structure of the stellar magnetic field, with the orientation of the
trapped fireball.  The presence of four X-ray `jets' requires that the
trapped fireball connect up with four bundles of magnetic field lines extending
to at least a few stellar radii.

\section{Braking via the internal exchange of angular momentum}

Now let us consider the exchange of angular momentum between the
the crustal superfluid neutrons and the rest of the magnetar.  
Because an SGR or AXP source is slowly rotating, $\Omega_{\rm cr} \sim 1$, 
the maximum angular velocity difference $\omega = 
\Omega_{\rm sf}-\Omega_{\rm cr}$ that can be maintained between superfluid and
crust is a much larger fraction of $\Omega_{\rm cr}$ than it is in an ordinary
radio pulsar -- and may even exceed it.  At the same time, these
sources are observed to spin down very rapidly, on a timescale
comparable to young radio pulsars such as Crab or Vela.  
If the rotation of the superfluid were to lag behind the crust in
the usual manner hypothesized for glitching radio pulsars, the maximum
glitch amplitude would increase in proportion to the spin period
(Thompson \& Duncan 1996, hereafter TD96; Heyl \& Hernquist 1999).
One deduces $\Delta P/P \simeq -1\times 10^{-5}$ by scaling to the
largest glitches of the Crab pulsar, and $\Delta P/P \simeq -1\times 10^{-4}$
by scaling to Vela.  

How would a glitch be triggered in a magnetar?  A sudden
fracture of the crust, driven by a magnetic field stronger than
$\sim 10^{14}$ G, induces a horizontal motion at the Alfv\'en
speed $V_{\rm A} = 1.3\times 10^{7}\,(B/10\,B_{\rm QED})\,(\rho/10^{14}\,{\rm g~
cm^{-3}})^{-1/2}$ cm s$^{-1}$, or higher.  This exceeds the
maximum velocity difference $V_{\rm sf}-V_{\rm cr}$
that can be sustained between superfluid and crust, before the
neutron vortex lines unpin (e.g. Link, Epstein, \& Baym 1993).  The internal
heat released in a large flare such as the August 27 event is
probably comparable to the external X-ray output, if the flare 
involves a propagating fracture of the neutron star crust.
This heat is $\sim 100$ times the minimum energy of $\sim 10^{42}$ erg
that will induce a sudden increase in the rate of thermal vortex creep
(Link \& Epstein 1993).  For both reasons, giant flares from magnetars
probably trigger the widespread unpinning of superfluid vortices in the
crust and hence large rotational glitches.  Magnetically-driven fractures 
have also been suggested as the trigger for vortex unpinning in ordinary radio
pulsars (Thompson \& Duncan 1993, hereafter TD93; 
Ruderman, Zhu, \& Chen 1998).

The observation of a period increase associated with the August 27
outburst leads us to re-examine whether the superfluid should, in 
fact, maintain a faster spin than the crust and charged interior
of the star.   Transport of superfluid vortices by thermal creep will
cause the angular velocity lag $\omega$ to relax to its equilibrium value
$\omega_\infty$ on a timescale
\be
t_r^{-1} = \left|{\partial\Omega_{\rm cr}\over\partial t}\right|
\left({\partial\ln V_{\rm cr}\over \partial\omega}\right)_{\omega_\infty},
\ee
if the creep is driven primarily by spindown (Alpar, Anderson, Pines, \&
Shaham 1984;  Link, Epstein, \& Baym 1993).  The partial derivative of the
creep velocity  $\partial V_{\rm cr}/\partial\omega$ depends mainly on 
temperature and density.  As a result, this relaxation time is expected
to be proportional to $t/\Omega_{\rm cr}$ at constant temperature.  Comparing
with a prompt (intermediate) relaxation time of $\sim 1$ day ($\sim 1$ week)
for glitches of the Crab pulsar ($t\simeq 10^3$ yr; Alpar et al.~1996), one
infers $t_r \sim 1$ ($10$) {\it years} for a magnetar of spin period $6$
s and characteristic age $P/\dot P = 3000$ yr.  

The response of the crust to the evolving magnetic field is expected
to be a combination of sudden fractures and plastic deformation.
When the temperature of the crust exceeds about $\sim 0.1$ of the melt
temperature, it will deform plastically (Ruderman 1991).  
One deduces  $T \simeq 2.4\times 10^8\,(B/10^2\,B_{\rm QED})^2$ K
for magnetars of age $\sim 10^4$ yr (TD96; Heyl \& Kulkarni 1998).
Plastic deformation is also expected when $B^2/4\pi > \mu$ in the deep crust
(TD96).   In a circumstance where the magnetic field is transported
through the stellar interior on a timescale shorter than the age of the
star, departures from corotation between superfluid and crust are primarily
due to advection of the superfluid vortices across the stellar surface by
the deforming crust, {\it not} due to spindown.  (Recall the principal
definition of a magnetar:  a neutron star in which magnetism, not rotation,
is the dominant source of free energy.)   If these deformations
occur on a timescale much less than the spindown age, they will control
the equilibrium lag between the rotation of the superfluid and crust.

Indeed, the SGR bursts provide clear evidence for deformations on 
short timescales.   More precisely, a large burst such as the August 27
event may be preceded (or followed) by an extended period of slow, plastic
deformation.  If the superfluid starts near corotation with the crust,
this process will take angular momentum out of the superfluid, and force
its rotation to lag behind the rest of the star.  A glitch
triggered by a violent disturbance such as the August 27 event will
then cause the neutron star crust to {\it spin down}.  

The angular momentum of the thin shell of crustal superfluid can be
expressed simply as
\be
J_{\rm sf} = {\kappa\over 2}M_{\rm sf} R_\star^2
\int_{-1}^1 d(\cos\theta)\,\cos^2\theta\,n_V(\theta),
\ee
when the cylindrical density $n_V(\theta)$ of neutron vortex lines depends
only on angle $\theta$ from the axis of rotation.  Here $\kappa =
h/2m_n$ is the quantum of circulation, and we neglect that the rotational
deformation of the star.  One observes
from this expression that the outward motion of vortex lines
reduces $J_{\rm sf}$, because the weighting factor $\cos^2\theta$ decreases
with distance from the axis of rotation.  

The simplest deformation of the neutron star crust, which preserves
its mass and volume, involves a rotational twist of a circular patch through
an angle $\Delta\phi$.  Indeed, the stable stratification of the star
(Reisenegger \& Goldreich 1992) forces the crust to move horizontally,
parallel to the local equipotential surfaces.  For this reason, one
can neglect horizontal displacements of the crustal material that
are compressible in the two non-radial dimensions.   The patch has radius
$a \ll R_\star$ and is centered at an angle $\theta$ from the axis of rotation.
The superfluid is assumed initially to corotate with the crust, 
$\Omega_{\rm sf} = \Omega_{\rm cr}$, everywhere within the patch, so that
$n_V(\theta) = 2\Omega_{\rm cr}/\kappa$.   As the patch is rotated, the
number of vortex lines per unit {\it surface area} of crust is conserved.
A piece of crust that moves from $\theta_i$ to $\theta_f$ ends up with a
vortex density  $n_V = (2\Omega_{\rm cr}/\kappa)\cos\theta_i/\cos\theta_f$. 
The vortex lines are squeezed together in a piece of the crust that moves
away from the rotation axis, and are spread apart if the movement is
in the opposite direction.  If the vortex density is smoothed
out in azimuth following this process, the net decrease in
the angular momentum of the superfluid is
\be\label{deltajsf}
{\Delta J_{\rm sf}\over J_{\rm sf}} = -{3\over 4}\left({a\over R_\star}\right)^4\,
\Bigl(1-\cos\phi\Bigr)\sin^2\theta.
\ee
Here, $J_{\rm sf} =  {2\over 3}M_{\rm sf}\Omega_{\rm cr}R_\star^2 \simeq
10^{-2}I_\star\Omega_{\rm cr}$ is the total angular momentum of the
crustal superfluid.

A transient, plastic deformation of the crust would induce a measurable
spinup of the crust, by forcing the neutron superfluid further from
corotation with the crust.  Such a gradual glitch would have the same
negative sign as in ordinary radio pulsars, but would not necessarily
involve any sudden unpinning of the vortex lines.  For example, rotation of
a patch of radius $a = {1\over 3} R_\star$ 
through an angle $\Delta\phi \sim 1$ radian would cause a period decrease
$\Delta P/P = \Delta J_{\rm sf}/(I_\star-I_{\rm sf})\Omega_\star 
= -4\times 10^{-5}$. A transient spinup of this magnitude may
have been observed in the AXP source 1E2259+586 (Baykal \& Swank 1996).
That excursion from a constant, long term spindown trend can be modelled
with a glitch of amplitude $\Delta P/P \simeq -3\times 10^{-5}$, although
the X-ray period observations are generally too sparse to provide a unique fit.

\section{The long-term spin-down of SGRs and AXPs}

Let us now consider the persistent spindown rate of SGR 1900+14, and its
broader implications for the ages and spindown histories of the SGR and 
AXP sources.  Recall that the spindown rate was almost constant at 
$\dot P \simeq 6.1\times 10^{-11}$ s/s before May 1998,
and after August 28 1998 (Paper I).   A May 1997 measurement of $P$
revealed a 5\% deviation from this trend; and larger
variations in the `instantaneous' spindown rate ($\sim 40$\%) were found 
by RXTE in September 1996 and May/June 1998.  

Another important constraint comes from the observed angular position 
of SGR 1900+14. \  It lies just 
outside the edge of the $\sim 10^4$ yr-old supernova remnant 
G42.8+0.6  (Hurley et al.~1994; Vasisht et al.~1994).  
A strong parallel can be drawn
with SGR 0526-66, which also emitted a giant flare (on 5 March 1979) and 
is projected to lie inside, but near the edge of, SNR N49 in the 
Large Magellanic Cloud (Cline et al.~1982).  
The other known SGRs also have positions coincident with supernova remnants of 
comparable ages (Kulkarni \& Frail 1993; Kulkarni et al.~1994;
Murakami et al.~1994; Woods et al.~1999b; Smith, Bradt, \& Levine 1999;
Hurley et al.~1999d). 
\ It seems very likely that these physical 
associations are real; so we will hereafter adopt the hypothesis that 
SGR 1900+14 formed at the center of SNR G42.8+0.6.  \  The implied 
transverse velocity is 
\be\label{vtrans}
V_{\perp} \approx 3400 \, \left({D\over 7 \, \hbox{kpc}} \right) 
\, \left({t\over 10^4 \, \hbox{yr}} \right)^{-1} \, \hbox{km s}^{-1}
\ee
(Hurley et al.~1996; Vasisht et al.~1996; Kouveliotou et al.~1999). 
Several mechanisms may 
impart large recoil velocities to newborn magnetars (Duncan \& Thompson 1992,
hereafter ``DT92"), but this very high speed indicates that 
an age much less than $1 \times 10^4$ yrs is unlikely.   

In this context,
the short charactersitic spindown age $P/2\dot P\sim 1400$ yr of 
SGR 1900+14 gives evidence that the star is currently in a 
transient phase of accelerated spindown (Kouveliotou et al.~1999).  
The almost identical
spindown age measured for SGR 1806-20 suggests that a similar effect
is being observed in that source (Kouveliotou et al.~1998; Table 1).  
If each SGR undergoes accelerated spindown during a minor
fraction $\epsilon_{\rm active}\sim P/\dot Pt_{\rm SNR}\sim 0.25$ of its
life, then its true age increases to 
\be\label{tactive}
t = \epsilon_{\rm active}^{-1}\,\left({P\over\dot P}\right).
\ee

\placetable{tbl-1}
\begin{noindent}
\begin{center}
\begin{deluxetable}{cccllcc}
\scriptsize
\tablecaption{SGRs and AXPs with Measured Spin-down Rates and Associations with 
Supernova Remnants  \label{tbl-1}}
\tablewidth{0pt}

\tablehead{
\colhead{Source}                      &
\colhead{Period}                      &
\colhead{Period Derivative}           &  
\colhead{t$_{\rm MDR}$}               &  
\colhead{t$_{\rm SNR}$}               &  
\colhead{t$_{\rm MDR}$/t$_{\rm SNR}$} \\  
\colhead{}                            &
\colhead{s}                           &
\colhead{s s$^{-1}$}                  &
\colhead{yr}                          &
\colhead{yr}                          &
\colhead{}
}

\startdata

SGR~$1806-20$\tablenotemark{a}   &   7.47  &  8.3 $\times$ 10$^{-11}$  &
1430    &  $\sim$ 10$^4$   &  $\sim$ 0.1  \nl 

SGR~$1900+14$\tablenotemark{b}   &   5.16  &  6.1 $\times$ 10$^{-11}$  &  
1340    &  $\sim$ 10$^4$   &  $\sim$ 0.1  \nl 

AXP~$1E2259+586$\tablenotemark{c}  &   6.98  &  5.0 $\times$ 10$^{-13}$  & 
220,000   &  $\sim$ 13,000    &  $\sim$ 10   \nl 

AXP~$1E1841-045$\tablenotemark{d}  &   11.8  & 4.13 $\times$ 10$^{-11}$  &  
4570    &  $\sim$ 2000     &  $\sim$ 2.3  \nl

\enddata

\tablenotetext{a}{Kouveliotou et al. 1998}
\tablenotetext{b}{This work and references herein}
\tablenotetext{c}{Mereghetti, Israel, \& Stella 1998 and references therein \\
		  Wang et al. 1992}
\tablenotetext{d}{Vasisht \& Gotthelf 1997 \\
		  Gotthelf \& Vasisht 1997 \\
		  Gotthelf, Vasisht, \& Dotani 1999}

\end{deluxetable}
\end{center}
\end{noindent}

\subsection{Wind-Aided Spindown}

Seismic activity will accelerate the  
spindown of an isolated neutron star, if the star is slowly rotating and
strongly magnetized (Thompson \& Blaes 1998, hereafter ``TB98").  
Fracturing in the crust generates seismic waves which 
couple directly to magnetospheric Alfv\'en modes and to the 
relativistic particles that support the associated currents. 
The fractures are frequent and low energy ($\sim 10^{35}$ erg)
when the magnetic field is forced across the crust by compressive
transport in the core (TD96).  When the persistent luminosity $L_{\rm A}$ of
waves and particles exceeds the magnetic dipole  luminosity $L_{\rm MDR}$ 
(as calculated from the stellar dipole field and angular velocity), 
the spindown torque increases by a factor $\sim \sqrt{L_{\rm A}/L_{\rm MDR}}$.

This result follows directly from our treatment of hydrodynamic torques
in \S 2.  Magnetic stresses force the relativistic wind to co-rotate
with the star out to the Alfv\'en radius $R_{\rm A}$, which is determined by
substituting $L_{\rm A}$ for $L_{\rm X}$ in eq. (\ref{Ralf}):
\be\label{raqui}
{R_{\rm A}\over R_\star} = 1.6\times 10^4\,L_{\rm A\,35}^{-1/4}\,
\left({B_\star\over 10\,B_{\rm QED}}\right)^{1/2}.
\ee
The torque then has the form $I \dot{\Omega} = - \Lambda (L/c^2) R_{\rm A}^2$, 
or equivalently
\be\label{pdot}
\dot{P} = \Lambda \, {B_\star R_\star^3 \over I_\star} \ 
\left({L_{\rm A} \over c^3}\right)^{1/2} \ P.
\ee
Here, $\Lambda$ is a numerical factor of order unity that depends on
the angle between the angular velocity ${\bf\Omega}$ and the dipole
magnetic moment ${\bf m}_\star$.
One finds $\Lambda \approx {2\over 3}$ by integrating eq. (\ref{domega})
over polar angle, under the assumption that ${\bf\Omega}$ and 
${\bf m}_\star$ are
aligned, that the ratio of mass flux to magnetic dipole flux is constant,
and that the magnetic field is swept into a radial configuration between
the Alfv\'en radius and the light cylinder.  This normalization is $\sim 6$
times larger than deduced by TB98 for a rotator
with ${\bf m}_\star$ inclined by 45$^\circ$ with respect to ${\bf \Omega}$:
they considered the enhanced torque resulting from the sweeping out of
magnetic field lines, but not the angular momentum of the outflow itself. 

The dipole magnetic field inferred from $P$ and $\dot P$ depends on the
persistent wind luminosity.  Normalizing $L_{\rm A}$ to
the persistent X-ray luminosity, $L_{\rm A} = L_{\rm A\,35}\times 10^{35}$ erg s$^{-1}$,
one finds for SGR 1900+14,
\be\label{bstar}
B_\star = 3 \times 10^{14} \ L_{\rm A\,35}^{-1/2} \ 
\left({\Lambda\over 2/3}\right)^{-1} \ I_{45} \
\left({\dot{P}\over 6\times10^{-11}}\right) \ 
\left({P\over 5.16 \, \hbox{s} }\right)^{-1}  \ \hbox{G.}
\ee
{\it A very strong magnetic field is needed to channel 
the flux of Alfv\'en waves and particles in co-rotation with the star out 
to a large radius.  This extended ``lever arm" enhances the magnetic 
braking torque for a given wind luminosity.}

The surface dipole field of SGR 1900+14 is inferred to be less than $B_{\rm QED} = 
4.4\times 10^{13}$ G only if $L_{\rm A} > 10^{37}$ erg s$^{-1}$.  That is,
the wind must be $\sim 30-100$ times more luminous than the time-averaged X-ray
output of the SGR in either quiescent or bursting modes.   Such
a large wind luminosity may conflict with observational bounds on the
quiescent radio emission of SGR 1900+14 (Vasisht et al.~1994; 
Frail, Kulkarni, \& Bloom 1999).  From these considerations alone
(which do not involve the additional strong constraints from bursting activity)
we find it difficutl to reconcile the observed spindown rate of SGR 1900+14
with dipole fields typical of ordinary radio pulsars (as suggested recently by
Marsden, Rothschild, \& Lingenfelter 1999).  

Note also that the synchrotron nebula surrounding SGR 1806-20
(Frail \& Kulkarni 1993), thought until recently to emanate from
the SGR itself and to require a particle source of luminosity
$\sim 10^{37}$ erg s$^{-1}$ (TD96), appears instead to be associated
with a nearby luminous blue variable star discovered by Van Kerkwijk
et al.~(1995).  The new IPN localization of the
SGR source (Hurley et al.~1999b) is displaced by 12$''$ from the peak
of the radio emission.  There is no detected peak in radio emission 
at the revised location.  Since the two SGRs have nearly identical
$\dot{P}/P$, we estimate a dipole field
$B_\star = 3\times 10^{14} \ L_{\rm A\,35}^{-1/2}$ G for SGR 1806-20.

During episodes of wind-aided spindown, the period grows exponentially:
\be
P(t) = {\cal P} \ \hbox{exp}(t/\tau_{\rm w}),
\ee
if the luminosity
$L_{\rm A}$ in outflowing Alfv\'en waves and relativistic particles remains constant. 
In this equation, $\tau_{\rm w} \equiv P/\dot{P} = I_\star c^{3/2}/
(\Lambda B_\star R_\star^3 L_{\rm A}^{1/2})$
is a characteristic braking time, and  ${\cal P}$ is the rotation period
at the onset of wind-aided spindown.
If $L_{\rm A}$ has remained unchanged over the lifetime of the star, then 
${\cal P}$ would be set by the
condition that the Alfv\'en radius sit inside the light cylinder,
${\cal P} = 2\pi ( B_\star^2 R_\star^6/c^3L_{\rm A})^{1/4} =
1.9 \, L_{\rm A\,35}^{-1/4} \, (B_{\star\,14}/3)^{1/2}$ s (cf.~eq.~[\ref{Ralf}]).
(Here, $B_\star = 10^{14} B_{\star\,14}$ G is the polar magnetic field.)

The narrow distribution of spin periods in the SGR/AXP
sources ($P = 5$---12 s) would be hard to explain if every source
underwent this kind of extended exponential spindown; but the 
possibility cannot be ruled out in any one source.  The total age of 
such a source would be 
\be\label{sdage}
t = (P/ \dot{P}) \ \ln(P/ {\cal P}) + t({\cal P}),
\ee
where $t({\cal P})$ is the time required to spin down to period 
${\cal P}$. \ Notice that $\dot{P}\propto P$ at constant $L_{\rm A}$, as
compared with $\dot{P}\propto P^{-1}$ in the case of magnetic dipole
radiation (MDR).   The net result is to {\it lengthen}
the spindown age deduced from a given set of $P$ and $\dot{P}$, relative to 
the usual estimate $t_{\rm MDR} \equiv P/2\dot{P}$ employed for radio pulsars.
Note also that $P/\dot P$ remains constant throughout episodes of
wind-aided spindown.

Applying these results to SGR 1900+14 (eq.~[\ref{bstar}]), we would 
infer that wind-aided spindown has been operating for 
$(P/\dot P)\ln(P/{\cal P}) = 2700$ yrs (assuming a steady wind 
of luminosity $L_{\rm A\,35} = 1$).  Its total age, including the age
$t({\cal P})$ at the onset of wind-aided braking, would be
$2700 + 1300 = 4000$ yrs.  (This number only increases to 5600 yrs if
$L_{\rm A}$ increases to $10^{36}$ erg s$^{-1}$.)  This age remains
uncomfortably short to allow a physical assocation with SNR G42.8+0.6: 
it would imply a transverse recoil velocity $V_{\perp} \approx 0.03 \, 
(D/ 7 \, \hbox{kpc})\,  c$  [eq.~(\ref{vtrans})]. 

The age of SGR 1900+14 can be much longer, and $V_\perp$ much
smaller, if the accelerated spindown we now observe
occurs only {\it intermittently} (eq.~[\ref{tactive}]).
In the magnetar model, it is plausible that small-scale seismic activity 
and Alfv\'en-driven winds are only vigorous during transient episodes, 
which overlap periods of bursting activity (\S 4.4 below).

\subsection{Connection with Anomalous X-ray Pulsars}

If each magnetar undergoes accelerated spindown only for a fraction 
$\epsilon_{\rm active}\sim P/\dot Pt_{\rm SNR}\sim 0.25$ of its life 
(eq.~[\ref{tactive}]),
then {\it the observed SGRs should be outnumbered some
$\epsilon_{\rm active}^{-1}\sim 4$ times by inactive sources 
that spin down at a rate $\dot P \leq P/2t_{\rm SNR}$}.

The Anomalous X-ray Pulsars (AXPs) have been identified as such inactive 
SGRs (Duncan \& Thompson 1996; TD96; Vasisht \& Gotthelf 1997;
Kouveliotou et al. 1998).  Although 
harder to find because they do not emit bright bursts, 6 AXPs are already 
known in our Galaxy, as compared with 3 Galactic SGRs.  
Table 1 summarizes the spin behavior and age estimates of the
two AXP sources that are presently associated with supernova remnants
(1E2259+586 and 1E1841-045).
Their characteristic ages are larger than those of SGRs 1900+14 and 1806-20.

The characteristic age of 1E2259+586 also appears to be much longer,
by about an order of magnitude, than the age of the associated
SNR CTB 109.  From Wang et al. (1992),
\be 
t_{\rm SNR} = 13,000 \, 
\left({E_{\rm SN}\over 0.4 \times 10^{51} \,\hbox{erg}}\right)^{-1/2} 
\, \left({n\over 0.13 \, \hbox{cm}^{-3}}\right)^{1/2} \, \hbox{yr,}
\ee 
where $E_{\rm SN}$ is the supernova energy and $n$ is the ISM particle density
into which the remnant has expanded.
Such a large characteristic age has a few possible 
explanations in the magnetar model.  First, the source may previously have
undergone a period of wind-aided spindown that increased its
period to $\sim 4$ times the value that it would have reached
by magnetic dipole braking alone.  Indeed, there is marginal
evidence for an extended X-ray halo surrounding the source, suggesting
recent output of energetic particles (Rho \& Petre 1997).  

Alternatively, the long characteristic age of 1E2259+586 could be caused 
by significant decay of the dipole field (TD93 \S 14.3 and 15.2); or by 
the alignment of a vacuum magnetic dipole with the axis of rotation
(Davis \& Goldstein 1970; Michel \& Goldwire 1970). 
Episodes of seismic activity can increase 
the spindown torque in aligned rotators both by driving the conduction
current above the displacement current in the outer magnetosphere, and by
carrying off angular momentum in particles and waves.  Indeed, the outer
boundary of the rigidly corotating magnetosphere, calculated by Melatos
(1997) to lie at a radius\footnote{When the displacement current
dominates the conduction current.}
 $R_{\rm mag}/R_\star = 1\times 10^3\,\gamma^{-1/5}
(B_\star/10^{14}~{\rm G})^{2/5}$, is contained well inside the speed of light
cylinder, $R_{\rm lc}/R_\star = 3\times 10^4\,(P/6~{\rm s})$. 
Here, $\gamma$ is the bulk Lorentz factor of the streaming charges.
There may be some tendency toward an initial alignment of
${\bf m}_\star$ and ${\bf\Omega}$ in rapidly rotating neutron stars
that support a large scale $\alpha$-$\Omega$ dynamo.  However, as we
argue in \S 4.3, rapid magnetic field decay will generically force
${\bf m}_\star$ out of alignment with ${\bf\Omega}$ and the principal
axes of the star.  

The remarkable AXP 1841--045 discovered by Vasisht \& Gotthelf (1997) is
only $\sim 2000$ yr old, as inferred from the age of the counterpart supernova
remnant (Gotthelf \& Vasisht 1997).  The ratio $t_{\rm MDR}/t_{\rm SNR}$ is
consistent with unity, in contrast with all other magnetar candidates that
have measured spindown and are associated with supernova remnants (Table 1).
Of these sources, AXP 1841--045 is also unique in failing to show measurable
variations in its spindown rate, X-ray luminosity, or X-ray pulse shape over
10 years (Gotthelf, Vasisht, \& Dotani 1999);  nor has it emitted any X-ray
bursts, or evinced any evidence for a particle outflow through a radio
synchrotron halo.  These facts reinforce the
hypothesis that {\it departures from simple magnetic dipole breaking 
are correlated with internal activity in a magnetar}, and suggest that
inactive phases can occur early in the life of a magnetar. 

\subsection{Free Precession in SGRs and AXPs}

Magnetic stresses will distort the shape of a magnetar (Melatos 1999).
The internal magnetic field generated by a post-collapse $\alpha$--$\Omega$
dynamo is probably dominated by a toroidal component (DT92; TD93).  
A field stronger than $\sim 100 \, B_{\rm QED}$ is transported through the
core and deep crust of the neutron star on a timescale short enough
for SGR activity (TD96).  Such a magnetar is initially {\it prolate},
with quadrupole moment $\epsilon=1\times 10^{-5}\,(B_{\rm in}/100~B_{\rm QED})^2$ 
(Bonazzola \& Gourgoulhon 1996).  Rapid field decay may cause the
magnetic moment ${\bf m}_\star$ to rotate away from the long principal
axis $\hat{\bf z}$ of the star, irrespective of any initial tendency for
these two axes to align.  The distortion of the rotating figure
of the star induced by the rigidity of the crust can be neglected 
when calculating the spin evolution of the star, as long as
$B > 10^{12}\,(P/{\rm 1~s})^{-1}$ G (Goldreich 1970). 

This hydromagnetic distortion gives rise to free precession on 
a timescale 
\be\label{taupr}
\tau_{\rm pr} = {2\pi\over \epsilon\Omega} = 2\times 10^{-2}\,
\left({B_{\rm in}\over 100~B_{\rm QED}}\right)^{-2}\,
\left({P\over 6~{\rm s}}\right)\;\;\;\;\;\;{\rm yr}.
\ee 
Even when the magnetosphere is loaded with plasma, the spindown torque
will depend on the angle between ${\bf m}_\star$ and the angular velocity
${\bf\Omega}$.  \ Free precession modulates
this angle when ${\bf m}_\star$ is canted with respect to the long
principal axis $\hat{\bf z}$, and so induces a periodic variation in the
spindown torque.  {\it Observation of free precession in an SGR or AXP
source would provide a direct measure of its total magnetic energy.}

How may free precession be excited?  In the case of a rigid vacuum dipole, 
free precession is damped by the radiation torque if the inclination between 
${\bf m}_\star$ and $\hat{\bf z}$ is less than
$55^\circ$ (Goldreich 1970).  \ At larger inclinations,
free precession is excited.  
In the more realistic case of a plasma-loaded magnetosphere, the rate at
which free precession is excited or damped by electromagnetic
and particle torques is, unfortunately, not yet known.   An additional,
{\it internal} excitation mechanism, which may be particularly effective
in an active SGR, involves rapid transport of the field in short,
intense bursts.  This is a likely consequence of energetic flares like
the March 5 or August 27 events, which probably have occurred $\sim 10^2$
times over the lifetimes of these sources.   If the principal axes
of the star are rearranged on a timescale less than $\tau_{\rm pr}$, then
${\bf\Omega}$ will not have time to realign with the principal axes
and precession is excited.  Only if the magnetic field is transported
on a timescale longer than $\tau_{\rm pr}$, will ${\bf\Omega}$
adiabatically track the principal axes.

An interesting alternative suggestion (Melatos 1999) is that 
forced radiative precession in a magnetar drives the bumpy
spindown of the AXP sources 1E2259+586 and 1E1048-593 on a timescale
of years.  When ${\bf m}_\star$ is not aligned with ${\bf\Omega}$,
the asymmetric inertia of the corotating magnetic field
induces a torque along ${\bf\Omega}\times{\bf m}_\star$ (Davis
\& Goldstein 1970).  This near-field torque acts on a timescale
$\tau_{\rm nf}$ that is $(\Omega R/c)$ times the electromagnetic braking time:
\be\label{taunf}
\tau_{\rm nf} \simeq 0.3\,
\left({B_\star\over 10~B_{\rm QED}}\right)^{-2}\,\left({P\over 6~{\rm s}}\right)
\;\;\;\;\;\;{\rm yr}.
\ee 
As long as $\tau_{\rm nf} < \tau_{\rm pr}$, this near-field torque
drives an anharmonic wobble of the neutron star; in particular, 
Melatos (1999) considers the case where $\tau_{\rm nf} \sim \tau_{\rm pr}$.
However, inspection of equations (\ref{taupr}) and (\ref{taunf}) suggests
instead that $\tau_{\rm pr} \ll \tau_{\rm nf}$, because the magnetic
energy is dominated by an internal toroidal component.  In this
case, the near-field torque averages to zero (Goldreich 1970).
Note also that this mechanism is predicated on an evacuated inner
magnetosphere, although the nonthermal spectra of SGRs and AXPs
indicate that this may not be a good approximation (Thompson 1999).  
The model has the virtue of making clear predictions of the future 
rotational evolution of the AXPs, which will be tested in coming years.

\subsection{Almost Constant Long-term Spindown}

We now address the near-uniformity of the long-term spindown rate of
SGR 1900+14, before and after the August 27 outburst 
(Woods et al. 1999a; Marsden, Rothschild \& Lingenfelter 1999; Paper I).  
It provides an important clue to any mechanism causing acceleration 
of the rate of spindown. 

There appears to be no measurable correlation between
bursting activity and long-term spindown rate (Paper I).  This observation
is consistent with the occurence of short, energetic bursts:  the
period increment caused by the release of a fixed amount of energy
is smaller for outbursts of short duration $\Delta t$, scaling
as $(\Delta t)^{1/2}$ (eq. [\ref{deltap}]).   The implied constancy
of the magnetic dipole moment is also consistent
with the energetic output of the August 27 burst:  only $\simeq 0.01\,
(E_{\rm Aug~27}/10^{44}~{\rm erg})\,(B_\star/10\,B_{\rm QED})^{-2}$ of
the exterior dipole energy need be expended to power the burst. 
Indeed, if the burst is powered by a large-scale magnetic instability,
one infers, from this argument alone,
that the dipole field cannot be much smaller than $10\,B_{\rm QED}$.

An additional clue comes from the bursting history of SGR 1806-20.  \ In that
source, the cumulative burst fluence grows with time, in a piecewise
linear manner (Palmer 1999).  This indicates that there exist many
quasi-independent active regions in the star, each of which 
expends a fraction $\sim 10^{-5}$ of the total energy budget.
The continuous output of waves and particles from the star is therefore 
the cumulative effect of many smaller regions.  Nonetheless, the long term
uniformity of $\dot P$ requires the rate of persistent seismic
activity in the crust to remain carefully regulated over a period
of years (or longer), even though the bursting activity is much more
intermittent.  

Persistent seismic activity is excited in a magnetar 
by the {\it compressive} mode of ambipolar diffusion of
the magnetic field through the core (TD96).  The resulting compressive
transport of the magnetic field through the crust requires
frequent, low energy ($E \sim 10^{35}$ erg) fractures of the crust induced
by the Hall term in the electrical conductivity.
The total energy released in magnetospheric particles has the same
magnitude as the heat conducted out from the core to the stellar surface.
The (orthogonal) rotational model of ambipolar diffusion will shear the crust.
It can induce much larger fractures that create optically thick
regions of hot $e^\pm$ plasma trapped by the stellar magnetic field (TD95).
The strong intermittency of SGR burst activity appears to be closely tied
to the energy distribution of SGR bursts, which is 
weighted toward the largest events (Cheng et al.~1996).  This suggests
that the rate of low-energy Hall fracturing will more uniform, being
modulated by longer term variations in the rate of ambipolar diffusion
through the neutron star core.

Nontheless, the modest variability observed in the short term measurements
of $\dot P$ (Paper I) must be accounted for.   Stochastic fluctuations in
the rate of small-scale crustal fractures provide a plausible mechanism. 
An alternative source of {\it periodic}, short-term variability involves
free precession in a magnetar whose dipole axis is tilted from the long
principal axis (\S 4.3).

Although angular momentum exchange with the crustal superfluid is a promising
mechanism to account for the $\Delta P/P\sim 10^{-4}$ period shift associated
with the August 27 event, it is less likely to dominate long-term variations
in the spindown rate.  An order of magnitude increase in the spindown rate
driven such exchange could persist only for a small fraction 
$\sim 10^{-1} I_{\rm sf}/I_\star \sim 10^{-3}$ of the star's life.  
Moreover, a gradual deformation of the neutron star crust by magnetic
stresses will remove angular momentum from the superfluid and decrease
the rate of spindown.

\section{Changes in the Persistent X-ray Flux and Lightcurve}

The persistent X-ray lightcurve of SGR 1900+14 measured following
the August 27 event (Kouveliotou et al.~1999; Murakami et al.~1999) appears
dramatically different from the pulse profile measured earlier:  indeed,
the profile measured following the burst activity of May/June 1998 
(Kouveliotou et al. 1999) is identical to that measured in April 1998
(Hurley et al. 1999c) and September 1996 (Marsden, Rothschild \& Lingenfelter
 1999).  Not only did the pulse-averaged luminosity increase by a factor 
2.3 between the 1998 April 30 and 1998 September 17/18 ASCA observations 
(Hurley et al.~1999c; Murakami et al.~1999),  but the lightcurve also
simplified into a single prominent pulse, from a multi-pulsed profile
before the August 27 flare.  The brighter, simplified lightcurve is suggestive
of enhanced dissipation in the active region of the outburst 
(Kouveliotou et al.~1999).  We now discuss the implications of this
observation for the dissipative mechanism that generates the persistent
X-rays, taking into account the additional constraints provided by the
period history of SGR 1900+14. 

\subsection{Magnetic Field Decay}

The X-ray output of a magnetar can be divided into two components (TD96):
thermal conduction to the surface, driven by heating in the core and
inner crust; and external Comptonization and particle bombardment powered
by persistent seismic activity in the star.  Both
mechanisms naturally generate $\sim 10^{35}$ erg s$^{-1}$ in
continuous output.   The appearence of a thermal pulse at the
surface of the neutron star will be delayed with respect to a deep fracture
or plastic rearrangement of the neutron star crust, by the thermal
conduction time of $\sim 1$ year (e.g. Van Riper, Epstein, \& Miller 1991).
By contrast, 
external heating will vary simultaneously with seismic activity in the star.
We have previously argued that if 1E2259+586 is a magnetar, then the
coordinated rise and fall of its {\it two} X-ray pulses (as observed by Ginga; 
Iwasawa et al.~1992) requires the thermal component of the 
X-ray emission to be powered, in part, by particle bombardment of
two connected magnetic poles (TD96, \S 4.2).

Neither internal heating, nor variability in the rate of persistent
seismic activity, appears able to provide a consistent explanation
for the variable lightcurve of SGR 1900+14.  
Deposition of $\sim 10^{44}$ erg of thermal energy in the deep crust,
of which a fraction $1-\epsilon$ is lost to neutrino radiation,
will lead to an increased surface X-ray output of $\sim 3\times 
10^{35}\,(\epsilon/0.1)$ erg s$^{-1}$. 
If, in addition, the heated deposited per unit mass
is constant with depth $z$ in the crust, then the heat per unit area scales
as $\sim z^4$; whereas the thermal conduction time varies weakly with
$z$ at densities above neutron drip (Van Riper et al.~1991).
The outward heat flux should, as a result, grow monotonically.
This conflicts with the appearance of the new pulse profile of 
SGR 1900+14 no later than one day after the August 27 event.   
By the same token, a significant increase in persistent seismic activity
-- at the rate needed to power the increased persistent luminosity 
 $L_{\rm X} \sim 1.5\times 10^{35}(D/7~{\rm kpc})^2$ erg s$^{-1}$
(Murakami et al.~1999) -- would induce a measurable change in the 
spindown rate that was not observed.  

The observations require instead a steady particle source that is
confined to the inner magnetosphere.  A large-scale deformation of
the crust of the neutron star, which likely occured during the August
27 outburst, must involve a horizontal twisting motion (\S 3).  
If this motion were driven by {\it internal} magnetic 
stresses,\footnote{A sudden unwinding of an external magnetic field
could release enough energy to power the March 5 (or August 27) event,
but it was argued in TD95 that the timescale $\sim R_\star/c \sim 10^{-4}$
s would be far too short to explain the width of the initial $\sim 0.2$
s hard spike.  A pulse broadened by a heavy matter loading would suffer
strong adiabatic losses and carry a much greater kinetic energy than
is observed in $\gamma$-rays.  Shearing of the external magnetic field requires
internal motions that will, in themselves, trigger a large outburst
by fracturing the crust.}
then the external magnetic field lines connected to the rotating patch
would be twisted with respect to their opposite footpoints
(which we assume to remain fixed in position).  We suppose that
the twist angle decreases smoothly from a value $\theta_{\rm max}$ at the
center of the patch to its boundary at radius $a$.  This means that a component
of the twist will remain even after magnetic reconnection eliminates
any tangential discontinuities in the external magnetic field resulting
from the motion.  The current carried by the twisted bundle
of magnetic field is 
\be
I \simeq {\theta_{\rm max} \Phi c\over 8\pi L},
\ee
where $\Phi = \pi a^2 B_\star$ is the magnetic flux carried by the bundle
and $L$ is its length.  

The surface of an AXP or SGR is hot enough ($T \sim 0.5$ keV) to feed
this current via thermionic emission of $Z < 12$ ions from one end
of the flux bundle, and electrons from the other end.
In magnetic fields stronger than $Z^3\alpha_{\rm em}^2 B_{\rm QED} = 4\times
10^{13}\,(Z/26)^3$ G, even iron is able to form long molecular chains.
The cohesion energy per atom is 
\be
{\Delta E\over Z^3\times 13.6\,{\rm eV}} 
= 1.52\,\left({B\over Z^3\alpha_{\rm em}^2\,B_{\rm QED}}\right)^{0.37}
-{7\over 24}\left[\ln\left({B\over Z^3\alpha_{\rm em}^2 B_{\rm QED}}
\right)\right]^2.
\ee
In this expression, the first term is the binding energy per atom in
the chain (Neuhauser, Koonin, \& Langanke 1987; Lai, Salpeter, 
\& Shapiro 1992), from which we subtract the binding energy of an isolated atom
(Lieb, Solovej, \& Yngvason, 1992).  Thermionic emission of ions is
effective above a surface temperature
\be    
T_{\rm thermionic} \simeq {\Delta E\over 30}.
\ee
Substituting $B = 10\,B_{\rm QED} = 4.4\times 10^{14}$ G, one finds
that $T_{\rm thermionic}$ remains well below 0.5 keV for $Z < 12$, but
grows rapidly at higher $Z$.  Thus, the surface of a magnetar should
be an effective thermionic emitter for a wide range of surface compositions.

We can now estimate the energy dissipated by the current flow.
The kinetic energy carried by ions of charge $Z$ and mass $A$ is 
\be
L_{\rm ion} = \left({A\over Z}\right) {Im_p\phi\over e} =
3\times 10^{35}\,{\theta_{\rm max}A\over Z}\,
\left({B_\star\over 10\,B_{\rm QED}}\right)\,\left({L\over R_\star}\right)^{-1}\,
\left({a\over 0.5~{\rm km}}\right)^2
\;\;\;\;\;\;{\rm erg~s^{-1}}.
\ee
Here, $\phi \simeq g_\star R_\star = GM_\star/R_\star$ 
is the gravitational potential that the charges have to climb along
the tube, and we assume $M_\star = 1.4\,M_\odot$, $R_\star = 10$ km.  
Note that the particle flow estimated here is large enough to break up
heavy nuclei even where the outflowing current has a positive sign:
electrons returning from the opposite magnetic footpoint are energetic
enough for electron-induced spallation to be effective
(e.g. Schaeffer, Reeves, \& Orland 1982).

On what timescale will this twist decay?  Each charge accumulates a
potential energy $A m_p g z$ a height $z$ above the surface of the neutron
star.  Equating this energy with the electrostatic energy released
along the magnetic field, one requires a longitudinal electric field
$E = Am_p g/Ze$.  The corresponding electrical conductivity is
\be
\sigma = {I\over\pi a^2 E} = \left({Z\theta_{\rm max}\over 8\pi A}\right)
\,{eBc\over m_p g_\star L},
\ee
and the ohmic decay time is
\be
t_{\rm ohmic} = {4\pi\sigma L^2\over c^2} = 
\left({Z\theta_{\rm max}\over 2A}\right) {eB_\star L\over m_p g_\star c}
= 300\,\left({Z\theta_{\rm max}\over A}\right)\,
\left({B_\star\over 10~B_{\rm QED}}\right)\,\left({L\over 10~{\rm km}}\right)
\;\;\;\;\;{\rm yr}.
\ee
This timescale agrees with that obtained by dividing the persistent
luminosity $L_{\rm ion}$ into the available energy of the twisted magnetic
field.  Further twisting of the field lines would prolong or shorten
the lifetime of the current flow.

A static twist in the surface magnetic field will not produce a measurable
increase in the torque because the current flow is contained
well inside the Alfv\'en radius (eq. [\ref{raqui}]).  The 
particles that carry the current lose their energy to Compton scattering
and surface impact on a timescale $\sim R_\star/c$ or shorter.  By contrast,
a persistent flux of low amplitude Alfv\'en waves into the magnetosphere 
causes the wave intensity to build up, until the wave luminosity transported
beyond the Alfv\'en radius balances the continuous output of the neutron star
(TB98).  Thus, the particle flow induced by a localized twist in the
magnetic field lines supplements the particle output associated with
persistent seismic activity occuring over the larger volume of the star.

\subsection{Evidence Against Persistent Accretion}

Direct evidence that the persistent X-ray output of
SGR 1900+14 is {\it not} powered by accretion comes from
measurements one day after the August 27 outburst (Kouveliotou
et al.~1999).  The increase in persistent $L_{\rm X}$ is not consistent
with a constant spindown torque, unless there was a substantial
change in the angular pattern of the emergent X-ray flux following
the burst.   In addition, the radiative momentum deposited by that
outburst on a surrounding accretion disk would more than suffice
to expel the disk material, out to a considerable distance from
the neutron star.  In such a circumstance, the time to re-establish
the accretion flow onto the neutron star, via inward viscous diffusion
from the inner boundary $R_{\rm in}$ of the remnant disk, would greatly exceed
one day.\footnote{This estimate of the viscous timescale
is conservative for two reasons:  First, if the binding energy of
the disk material were balanced with the incident radiative energy, the inner
boundary of the remnant disk would like at even larger radius.  Second,
the central X-ray source may puff up the disk, which increases
$\tau_{\rm visc}$ (eq. [\ref{tvisc}]).}
Let us consider this point in more detail.

The accretion rate (assumed steady and independent of radius before the
outburst) is related to the surface mass density $\Sigma(R)$ of the 
hypothetical disk via
\be
\dot M = {2\pi R^2 \Sigma(R) \over t_{\rm visc}(R)}.
\ee
The viscous timescale is, as usual,
\be
t_{\rm visc}(R) \simeq \alpha_{\rm SS}^{-1}\,\left({H(R)\over R}\right)^{-2}\,
\left({R^3\over GM_\star}\right)^{1/2},
\ee
where $H(R)$ is the half-thickness of the disk at radius $R$ and
$\alpha_{\rm SS} < 1$ is the viscosity coefficient (Shakura \&
Sunyaev 1973).  Balancing the radiative momentum incident on a
solid angle $\sim 2\pi (2H/R)$ against the momentum 
$\sim \pi \Sigma(R) R^2(2GM_\star/R)^{1/2}$ of the disk material
moving at the escape speed, and equating the persistent X-ray luminosity
$L_{\rm X}$ with $GM_\star\dot M/R_\star$, one finds
\be\label{tvisc}
t_{\rm visc} = {E_{\rm Aug~27}\over L_{\rm X}}\,
\left({2GM_\star\over R_\star c^2}\right)^{1/2}\,
\left({R_{\rm in}\over R_\star}\right)^{1/2}\,
\left({H(R_{\rm in})\over R_{\rm in}}
\right).
\ee

The most important factor in this expression is the ratio of
burst energy to persistent X-ray luminosity, $E_{\rm Aug~27}/L_{\rm X} =
30~(E_{\rm Aug~27}/10^{44}~{\rm erg})\,(L_{\rm X}/10^{35}~{\rm erg~s^{-1}})^{-1}$
yr.  The timescale is long as the result of the enormous energy of the
August 27 flare, and the relatively weak persistent X-ray flux preceding it.
It is interesting to compare
with Type II X-ray bursts from the Rapid Burster and GRO J1744-28, which
are observed to be followed by dips in the persistent emission 
(Lubin et al.~1992; Kommers et al.~1997).  These bursts, which certainly
are powered by accretion, involve energies $\sim 10^4$ times smaller
and a persistent source luminosity that is $10^2-10^3$ times higher.
Indeed, the dips in the persistent emission
following the Type II bursts last for only 100-200 s, consistent with
the above formula.

Now let us evaluate eq. (\ref{tvisc}) in more detail.
At a fixed $\dot M$, the surface mass density of the disk increases
with decreasing $\alpha_{\rm SS}$, and so a conservative upper bound
on $t_{\rm visc}$ is obtained by choosing $\alpha_{\rm SS}$ to be small.
(Note that eq. (\ref{tvisc}) depends implicity on $\alpha_{\rm SS}$ only
through the factor of $R_{\rm in}^{1/2} \propto \alpha_{\rm SS}^{1/2}$.)
For the observed parameters $E_{\rm Aug~27} \simeq 10^{44}$ erg (Mazets 
et al.~1999) and $L_{\rm X} = 10^{35}$ erg s$^{-1}$ (before the August 27 outburst;
Hurley et al.~1999a), one finds $R_{\rm in} = 1\times 10^{10}$ cm when
$\alpha_{\rm SS} = 0.01$.  The corresponding thickness of the gas-pressure
dominated disk is (Novikov \& Thorne 1973) $H(R_{\rm in})/R_{\rm in} 
\simeq 5\times 10^{-3}$.  The timescale over which the persistent
X-ray flux would be re-established is extremely long, 
$t_{\rm visc} \simeq 10$ yr.  

One final note on disk accretion.  There is no observational evidence 
for a binary companion to any SGR or AXP (Kouveliotou 1999).  
Because of its large recoil velocity (eq.~[\ref{vtrans}]),
SGR 1900+14 almost certainly could not remain bound in a binary system. 
A similar argument applies to the other giant flare source, SGR 0526--66
(DT92).  Thus, any accretion onto SGR 1900~+14 would have to come 
from a fossil disk.  To remain bound, the initial radius of such a disk must
be less than $GM_\star/V_{\rm rec}^2 \sim 10^4$ km, for stellar recoil 
velocity $V_{\rm rec} \sim (3/2)^{1/2} V_\perp$  [eq.~(\ref{vtrans})].  
The behavior of a passively spreading remnant disk appears inconsistent
with the measured spin evolution of the AXP and SGR sources (Li 1999).

A trigger involving sudden accretion of an unbound planetesimal
(Colgate and Petschek 1981) is not consistent with the log-normal
distribution of waiting periods between bursts (Hurley et al.~1994) in 
SGR 1806-20.  An internal energy source is also indicated by the power-law
distribution of burst energies, with index $dN/dE \sim E^{-1.6}$ similar to the
Gutenburg-Richter law for earthquakes (Cheng et al.~1996). 
In addition, the mass of the accreted planetesimals 
must exceed $\sim 1/30$ times the mass of the Earth's Moon in the case
of the March 5 and August 27 events.  It is very difficult to understand
how the accretion of a baryon-rich object could induce a fireball as
clean as the initial spike of these giant flares (TD95, \S 7.3.1).  
When $B_\star \ll 10^{14}$ G, only
a tiny fraction $(B_\star/B_E)^2$ of the hydrostatic released
would be converted to magnetic energy;  here, $B_E \sim 10^{14}$ G
is the minimum field needed to directly power the outburst.

\section{Conclusions}

The observation (Paper I) of a rapid spindown associated with the August
27 event, $\Delta P/P = +1\times 10^{-4}$, provides an important
clue to the nature of SGR 1900+14.
We have described two mechanisms that could induce such a rapid loss of
angular momentum from the crust and charged interior of the star.
The torque imparted by a relativistic outflow during the
August 27 event is proportional to $B_\star$, but falls short by
an order of magnitude even if $B_\star \sim 10\,B_{\rm QED} = 4.4\times 10^{14}$ G.
Only if SGR 1900+14 released an additional $\sim 10^{44}$ erg for an 
extended period $\sim 10^4$ s immediately following the August 27 outburst
would the loss of angular momentum be sufficient.   (The integrated
torque increases with the duration $\Delta t$ of the outflow as
$(\Delta t)^{1/2}$; eq. [\ref{deltap}].)  

The alternative model, which we favor, involves a glitch driven
by the violent disruption of the August 27 event.  The unpinned neutron
superfluid will absorb angular momentum if it starts out spinning more
slowly than the rest of the star -- the opposite of the situation
encountered in glitching radio pulsars.  We have argued that a slowly
spinning neutron superfluid is the natural consequence of magnetic
stresses acting on the neutron star crust.  A gradual,  plastic
deformation of the crust during the years preceding the recent onset
of bursting activity in SGR 1900+14 would move the superfluid out of
co-rotation with the rest of the star, and slow its rotation.
The magnitude of the August 27 glitch can be crudely estimated by 
scaling to the largest glitches of  young, active pulsars with similar
spindown ages and internal temperatures.  Depending on the object
considered, one deduces $|\Delta P|/P \sim 10^{-5}-10^{-4}$.

This model for the August 27 period increment has interesting implications
for the longer-term spindown history of the Soft Gamma Repeaters
and Anomalous X-ray Pulsars.  It suggests that these objects
can potentially glitch, with or without associated bursts, and that
$P$ will suddenly shift {\it upward,} rather than downward as in radio 
pulsar glitches.  By the same token,
an accelerated rate of plastic deformation within a patch of the neutron star
crust will force the superfluid further out of co-rotation and induce a
transient (but potentially resolvable)
{\it spin-up} of the crust (TD96).  The magnitude of such a `plastic
spin-up' event (eq. [\ref{deltajsf}]), 
could approach that inferred for the August 27 event, but
with the usual (negative) sign observed in radio pulsar glitches.
Indeed, RXTE spin measurements provide evidence for a 
rapid spin-up of the AXP source 1E2259+586 (Baykal \& Swank 1996),
to the tune of $\Delta P/P = -3\times 10^{-5}$.   Transient variations
in the persistent X-ray flux of the AXP 1E2259+586, which were not
associated with any large outbursts, also require transient plastic
deformations of the neutron star crust (TD96).

The rapid spindown rate of SGR 1900+14 during the past few years, 
$\dot P = 6\times 10^{-11}$ s/s,
indicates that this SGR is a transient phase of {\it accelerated
spindown}, with stronger braking torques than would be produced by 
simple magnetic dipole radiation (Kouveliotou et al.~1999).  
Such accelerated spindown can be driven by magnetically-induced seismic
activity, with small-scale fractures powering a steady
relativistic outflow of magnetic vibrations and particles.  This outflow,
when channeled by the dipole magnetic field, carries away the star's 
angular momentum.  A very strong field, $B_\star \gg B_{\rm QED}$, is required
to give a sufficiently large ``lever arm" to the outflow.

Further evidence for episodic accelerated spindown comes from 
the two AXPs that are directly associated with supernova remnants:  
1E2259+586 and 1E1841-045 (\S 4.2). \
The characteristic ages $P/2 \dot P$ of these stars are  {\it longer}
than the the ages of the associated supernova remnant, and also longer than
the characteristic ages of the SGRs. This suggests that the AXPs are
magnetars observed during phases of seismic inactivity.

The constancy of the long-term spindown rate before and after the bursts
and giant flare of 1998 (Woods et al. 1999a; Marsden, Rothschild 
\& Lingenfelter 1999; Paper I)
gives evidence that the spindown rate correlates only weakly
with bursting activity.  It is easy to understand why short, intense 
bursts are not effective at spinning down a magnetar: 
the Alfv\'en radius (the length of the ``lever arm") decreases 
as the flux of Alfv\'en waves and particles increases.  

A persistent output of waves and particles
could be driven by the compressive mode of ambipolar diffusion
in the liquid neutron star interior (TD96).  As the magnetic field is forced
through the crust, the Hall term in the electrical conductivity induces
many frequent, small fractures ($\Delta E \sim 10^{35}$ erg).  By
contrast, large fractures of the crust are driven by shear stresses
that involve the orthogonal (rotational) mode of ambipolar
diffusion.  The greater intermittency of bursting activity is
a direct consequence of the dominance of the total burst fluence
by the largest bursts (Cheng et al.~1996).

Forced radiative precession could cause a short-term 
modulation of the spindown rate in a magnetar (Melatos 1999), but this
requires an evacuated magnetosphere that may not be consistent with the
observed non-thermal spectra of the SGR and AXP sources (Thompson 1999).
We have argued that transport of the neutron star's 
magnetic field will deform the principal axes of the star
and induce {\it free} precession.  The resulting modulation of the
spindown torque has an even shorter timescale (eq. [\ref{taupr}]),
and is potentially detectable.  

A twist in the exterior magnetic field induced by a large scale
fracture of the crust will force a persistent thermionic current through
the magnetosphere (\S 5).  The resulting steady output in particles
would explain the factor $\sim 2.3$ increase in the
persistent X-ray flux of SGR 1900+14 immediately
following the August 27 event (Murakami et al.~1999) if
$B_\star \sim 10B_{\rm QED}$ and the twist is through $\sim 1$ radian.
In this model, the simplification of the lightcurve -- into a single
large pulse -- is due to concentrated particle heating at the site
of the August 27 event.

We conclude by emphasizing the diagnostic potential of coordinated
measurements of spectrum, flux, bursting behavior and period derivative.
When considered together, they constrain not only the internal
mechanism driving the accelerated spindown of an SGR source, but also
the mechanism powering its persistent X-ray output.  For example:
an increase in surface X-ray flux will be delayed by $\sim 1$ year
with respect to an episode of deep heating (e.g. Van Riper et al.~1991); 
whereas a shearing and twisting of the external magnetic field of the neutron
star will drive a simultaneous increase in the rate of external particle
heating (TD96).  The magnetar model offers a promising framework in which
to interpret these observations.

\acknowledgments{\noindent We acknowledge support from
NASA grant NAG 5-3100 and the Alfred P. Sloan foundation (C.T.);
NASA grant NAG5-8381 and Texas Advanced Research Project grant ARP-028
(R.C.D.); the cooperative agreement NCC 8-65 (P.M.W.); and NASA grants
NAG5-3674 and NAG5-7808 (J.vP.). C.T. thanks A. Alpar and M. Ruderman for
conversations.}

\end{document}